\documentstyle[prl,aps,preprint]{revtex}
\begin{document}

\title{Spin-wave condensation and quantum melting of long-range
antiferromagnetic order in $t-J$ model.}

\author {O. P. Sushkov$^{a}$}
\address { School of Physics, The University of New South Wales
 Sydney 2052, Australia}

\maketitle

\begin{abstract}

Ground state wave function of two-dimensional $t-J$ model is found
at doping close to half filling. It is shown that the condensation
of Cooper pairs (superconducting pairing of mobile holes) and
the condensation of spin-waves into spin-liquid state are closely connected.
The effective spectrum of $S=1$ excitations,
spin-wave gap pseudo-gap $\Delta_M$, magnetic correlation length $\xi_M$, and
static magnetic formfactor $S_M(\bf q)$ are calculated.
\end{abstract}

\vspace{0.5cm}
\hspace{3.cm}PACS numbers: 71.27.+a, 74.20.Hi, 75.50.Ee
\vspace{1.cm}

\section{Introduction}
In the recent papers \cite{Flam94,Bel95} in the framework of $t-J$ model
we have demonstrated very strong  d-wave pairing between dressed quasiholes
induced by   spin-wave exchange. The pairing gives critical
temperature in a reasonable agreement with experimental data for copper
oxide superconductors.
In general terms our approach follows the studies, of Monthoux, Pines,
Scarlapino, Schriffer, and others, supporting idea of the magnetic
fluctuations mechanism of pairing\cite{Sch8,Bic9,Kam1,Bul1,Mon1,Mon2}.
For calculations we have used BCS-like
approximation for dressed quasiholes.  A similar approach to the
pairing of dressed quasiholes has been used by Dagotto, Nazarenko, and
Moreo \cite{Naz}. The important difference is that in our
work\cite{Flam94,Bel95}
the hole-hole interaction is derived from the parameters of the
t-J model, while in \cite{Naz} it is introduced  {\it ad hoc}
with magnitude adjusted to fit experimental data.

We consider the holes
as well as spin-waves moving in the background with long-range
antiferromagnetic (AF) order.  It means that we assume that the AF order is
preserved at distances smaller than inverse Fermi momentum $r \le 1/p_F$.
It is widely believed that this assumption is correct and present
work confirms this point. It is also widely believed that correct
spin structure of the ground state should be of the type of Anderson
spin-liquid state\cite{And}. For $t-J$ model
the instability of long-range AF order together with normal-hole
Fermi liquid was probably first pointed out by Shraiman and
Siggia\cite{Shr9}.
Later it has been demonstrated in numerous works (e.g. see  Refs.
\cite{Dom0,Sin0,Ede1,Iga2,Sus3}). It is due to the strong
interaction of spin-waves with mobile holes. However, a structure of the
ground state as well as a spectrum of excitations was not understood, and
spin-liquid state was not derived.

Now we understand that we have to consider a hole-Fermi-liquid
with superconducting pairing which interacts with long-range fluctuations of
AF ordering. Physical difference of the paired Fermi-liquid from normal one
is stiffness of the former.
In the present work we demonstrate that the long-range AF order is
also destroyed, but pairing is preserved. We derive explicitly the
spin-liquid state of the system. We show that selfconsistent translation
invariant spin-liquid solution exists at doping $\delta \ge \delta_c
\approx 3-4\%$. Using
this solution we calculate spin-wave pseudo-gap $\Delta_M$, magnetic
correlation length $\xi_M$ and
static magnetic formfactor $S(\bf q)$. In general terms the results which we
derive here in the framework of $t-J$ model are very close to the
picture of magnetic state discussed in the works of
Chubukov, Pines, Sachdev, Sokol, and Ye\cite{Chub93,Sok93,Sach94,Chub94}.

The paper has the following structure. In Sec.II we present the effective
Hamiltonian of $t-J$ model and discuss pairing of holes. In Sec.III
we formulate the idea of our approach presenting the simplified calculation
of spin-wave Green's function. Sec.IV presents calculation of spin-wave
polarization operator. In Sec.V spin-wave Green's function is calculated.
We also discuss the pseudo-gap in spin-wave spectrum. Sec.VI presents
calculations of magnetic correlation length $\xi_M$ and
static magnetic formfactor $S_M(\bf q)$. Finally our conclusions are
given in Sec. VII.

\section{Effective Hamiltonian and pairing of dressed holes}
The $t$-$J$ model is defined by the Hamiltonian
\begin{equation} \label{H}
  H = H_t + H_J
    = -t \sum_{<nm>\sigma} ( d_{n\sigma}^{\dag} d_{m\sigma} + \mbox{H.c.} )
     + J \sum_{<nm>}  {\bf S}_n \cdot {\bf S}_m,
\end{equation}
where $d_{n\sigma}^{\dag}$ is the creation operator of a hole with spin
 $\sigma$ ($\sigma= \uparrow, \downarrow$) at site $n$ on a
 two-dimensional square lattice.
The $d_{n\sigma}^{\dag}$ operators act in the Hilbert space
 with no double electron occupancy.
The spin operator is ${\bf S}_n = {1 \over 2} d_{n \alpha}^{\dag}
 ${\boldmath $\sigma$}$_{\alpha \beta} d_{n \beta}$.
$<nm>$ are the nearest-neighbour sites on the lattice.
Below we set $J$ as well as lattice spacing equal to unity.

At half-filling (one hole per site) $t$-$J$ model is equivalent to
the Heisenberg AF model\cite{Hir5,Gro7} which has long-range
AF order in the ground state\cite{Oit1,Hus8}. Let us denote the
wave function of this ground state by $|0\rangle$.
This is undoped system.
We consider the doped system basing on the ground state
of undoped one.
In spite of destruction of the long-range AF order it is convenient to use
$|0\rangle$
and corresponding quasiparticle excitations as a basis set in the problem
with doping.
The effective Hamiltonian for $t-J$ model in terms of these
quasiparticles was derived in the papers\cite{Suhf,Cher3,Kuch3}
\begin{equation}
\label{Heff}
H_{eff}=\sum_{{\bf k}\sigma}\epsilon_{\bf k}h_{{\bf k}\sigma}^{\dag}
h_{{\bf k}\sigma}+
\sum_{\bf q}\omega_{\bf q}(\alpha_{\bf q}^{\dag}\alpha_{\bf q}
+\beta_{\bf q}^{\dag}\beta_{\bf q}) +H_{h,sw} + H_{hh}.
\end{equation}
It is expressed in terms of usual spin-waves on AF
background $\alpha_{{\bf q}}, \beta_{{\bf q}}$
(see e.g. review\cite{Manousakis}), and
composite hole operators $h_{{\bf k}\sigma}$ ($\sigma = \pm 1/2$).
The summations over ${\bf k}$ and ${\bf q}$ are restricted inside the
Brillouin zone of one sublattice where
$\gamma_{\bf q}= {1\over 2} (\cos q_x+\cos q_y)\ge 0$.
Spin-wave dispersion is
\begin{equation}
\label{swdisp}
\omega_{\bf q}=2\sqrt{1-\gamma_{\bf q}^2}, \ \ \ \
\omega_{\bf q} \approx \sqrt{2}|{\bf q}|, \ at \ q \ll 1.
\end{equation}

Single hole properties are well established (for a review see
Ref.\cite{Dag4}). Wave function of a single hole can be represented as
 $\psi_{{\bf k}\sigma} = h_{{\bf k}\sigma}^{\dag} |0\rangle$.
At large $t$ the composite hole operator $h_{{\bf k}\sigma}^{\dag}$ has
complex structure. For example at $t/J=3$ the weight of bare hole in
$\psi_{{\bf k} \sigma}$ is  about 25\%, the weight of configurations
``bare hole + 1 magnon'' is  $\sim$50\%, and of configurations
``bare hole + 2 or more  magnons'' $\sim$25\%. Dressed hole is a normal
fermion. Hole energy $\epsilon_{\bf k}$
has minima at ${\bf k}={\bf k}_0$, where
${\bf k}_0=(\pm \pi/2, \pm\pi/2)$. For $t \le 5$ the dispersion can be
well approximated by the expression\cite{Sus2}
\begin{equation}
\label{hdisp}
\epsilon_{\bf k} \approx E_0+2-\sqrt{0.66^2+4.56t^2-2.8t^2\gamma_{\bf k}^2}
   + {1\over4} \beta_2 (\cos k_x -\cos k_y)^2.
\end{equation}
The decimals in this formula are some combinations of the
Heisenberg model spin correlators. Constant $E_0$ defines reference
level for the energy. To find $\epsilon_{\bf k}$ with respect to
undoped system one has to set $E_0=0$. However for present work it
is convenient to set $\epsilon_{\bf k_0}=0$, and therefore
$E_0=\sqrt{0.66^2+4.56t^2}-2$.
The coefficient $\beta_2$ is small and therefore the dispersion is almost
degenerate along the face of the magnetic Brillouin zone $\gamma_{\bf k}=0$.
According to Refs. \cite{Mart1,Giam3} $\beta_2 \approx 0.1\cdot t$
at $t \ge 0.33$.
To avoid misunderstanding we note that formula (\ref{hdisp}) certainly
is not valid for very large $t$ where the hole band width is saturated
at the value of the order of unity and does not increase with $t$.
However physically we are interested in $t \approx 3$ (e.g., see
Refs.\cite{Esk0,Fla1,BCh3}) where (\ref{hdisp}) works well.
Near the band minima ${\bf k}_0$ the dispersion (\ref{hdisp}) can be
presented in the usual quadratic form
\begin{equation}
\label{hdisp1}
\epsilon_{\bf p} \approx {1\over2} \beta_1 p_1^2 + {1\over2} \beta_2 p_2^2,
 \hspace{1.0cm} \beta_2 \ll \beta_1,
\end{equation}
where $p_1$ ($p_2$) is the projection of ${\bf k}-{\bf k}_0$
 on the direction orthogonal (parallel) to the face of the
 magnetic Brillouin zone (Fig.1). From Eq.(\ref{hdisp}) for $t \gg 0.33$
we find $\beta_1 \approx 0.65 t$, hence the mass anisotropy is
$\beta_1/\beta_2 \approx 7$.

The interaction of a composite hole with  spin waves
 is of the form (see, e.g., Refs.\cite{Mart1,Liu2,Suhf})
\begin{eqnarray}
\label{hsw}
&& H_{h,sw} = \sum_{\bf k,q} g_{\bf k,q}
    \biggl(
  h_{{\bf k}+{\bf q}\downarrow}^{\dag} h_{{\bf k}\uparrow} \alpha_{\bf q}
  + h_{{\bf k}+{\bf q}\uparrow}^{\dag} h_{{\bf k}\downarrow} \beta_{\bf q}
  + \mbox{H.c.}   \biggr),\\
&& g_{\bf k,q} = 2\sqrt{2}f
 (\gamma_{\bf k} U_{\bf q} + \gamma_{{\bf k}+{\bf q}} V_{\bf q}),\nonumber
\end{eqnarray}
where
$U_{\bf q}=\sqrt{{1\over{\omega_{\bf q}}}+{1\over 2}}$ and
$V_{\bf q}=-sign(\gamma_{\bf q})\sqrt{{1\over{\omega_{\bf q}}}-{1\over 2}}$
are the parameters of Bogoliubov transformation diagonalizing spin-wave
Hamiltonian. The hole spin-wave coupling constant $f$ is a function of $t$
evaluated in the work\cite{Suhf}.
For large $t$ the coupling constant is $t$-independent $f \approx 2$.
Let us stress that even for $t > J$ the quasihole-spin-wave
interaction (\ref{hsw}) has the same form as
for $t \ll J$ ({\em i.e.} as for bare hole  operators)
with an added renormalization factor (of the order $J/t$ for $t \gg J$).
This remarkable property of the $t$-$J$ model is due to the absence
 of single loop correction to the hole-spin-wave vertex.
It was first stated implicitly by Kane, Lee, and Read\cite{Kan9}.
In Refs.\cite{Mart1,Liu2,Suhf} it was
explicitly demonstrated that the vertex corrections with different kinematic
structure are of the order of few percent at $t/J \approx 3$.
There is also some $q$-dependence of the coupling constant $f$.
For example $f(q=\pi)\approx 1.15f(q=0)$ at $t/J = 3$ (see Refs.
\cite{Kuch3,Bel95}). However this dependence is weak, it is practically
beyond the accuracy of the calculation of the renormalized value of $f$.
Therefore we neglect this dependence.

Finally there is contact hole-hole interaction $H_{hh}$ in the
effective Hamiltonian (\ref{Heff}). $H_{hh}$ is discussed in details in
the Refs.\cite{Cher3,Kuch3,Bel95}. It is proportional to some function
$A(t)$. For small $t$ this function approaches the value  -0.25, which
gives the well known hole-hole attraction induced by the reduction of the
number of missing antiferromagnetic links. However for
realistic superconductors $t\approx 3$ (see e.g.Refs.\cite{Esk0,Fla1,BCh3}).
Surprisingly function $A(t)$ vanishes exactly at $t\approx 3$, and it
means that the mechanism of contact hole-hole attraction is switched off.
In contrast the spin-wave exchange mechanism $H_{h,sw}$ is negligible
for small $t$ where $f \sim t$, and it is most important at
large $t$ where $f$ approaches 2. We are interested
in ``physical'' values of $t$: $t \approx 3$. Therefore in the present
work we neglect contact interaction ($H_{hh}=0$) and consider only
the hole-spin-wave interaction at $t=3$. Corresponding value of $f$
according to\cite{Suhf} is $f=1.8$.

The interaction which gives hole-hole superconducting pairing arises from
spin-wave exchange (Fig.2). From (\ref{Heff}) we find
\begin{equation}
\label{Vkk}
V_{\bf k, k'} = -2 {
      { g_{\bf k, q} g_{\bf k', -q} }
   \over
     { -\omega_{\bf q} - E_{\bf k} - E_{\bf k'} }},
\end{equation}
where ${\bf q}={\bf k}+{\bf k'}$, and $E_{\bf k}$ is the energy
of quasihole excitation above the BCS ground state (see below).
We consider the interaction in static approximation.
Justification of this approximation is as follows.
At small hole concentration $\delta \ll 1$, the holes are localized in momentum
space in the vicinity of the minima of the band
${\bf k}_0 = (\pm \pi/2, \pm \pi/2)$ and Fermi surface consists of ellipses.
The Fermi energy and Fermi momentum of non-interacting holes are
\begin{equation} \label{ep}
 \epsilon_F \approx \frac{1}{2} \pi (\beta_1 \beta_2)^{1/2} \delta, \qquad p_F
\approx \sqrt{p_{1F}p_{2F}} \approx (\pi \delta)^{1/2}.
\end{equation}
The Fermi momentum $p_F$ is measured from the center of corresponding ellipse.
In pairing, the exchange of spin-waves with typical momenta $\pi > q > p_F$
is the most important (see also Refs.\cite{Flam94,Bel95,Kuch3}). The energy of
such spin wave is much higher than the typical energy of a Cooper pair
\begin{equation}
\label{omega}
\omega_q \sim q > p_F \sim (\pi \delta)^{1/2} \gg \epsilon_F \sim (\beta_1
\beta_2)^{1/2} \delta.
\end{equation}
This justifies the static approximation (\ref{Vkk}) for hole-hole interaction
(see also comment at end of section V). This is much different from usual
phonon induced pairing where Debye's frequency is much lower than the Fermi
energy.

With trial function of ground state
\begin{equation}
\label{Psi}
 | \Psi \rangle = \prod_{\bf k} (u_{\bf k}+v_{\bf k}
   h^{\dag}_{{\bf k}\uparrow} h^{\dag}_{-{\bf k}\downarrow} ) |0 \rangle.
\end{equation}
we get conventional BCS equation for superconducting gap $\Delta_{\bf k}$
\begin{eqnarray}
\label{BCS}
&&\Delta_{\bf k}= -\frac{1}{2} \sum_{\bf k'} V_{\bf kk'}
  \frac{\Delta_{\bf k'}}{E_{\bf k'}} \tanh {{E_{\bf k'}}\over{2T}},\\
&&E_{\bf k} = \sqrt{\xi_{\bf k}^2+ \Delta_{\bf k}^2},\nonumber\\
&&u_{\bf k}=\sqrt{{1\over{2}}(1+ \xi_{\bf k}/E_{\bf k})},\nonumber\\
&&v_{\bf k}=sign{\Delta_{\bf k}} \cdot
\sqrt{{1\over{2}}(1- \xi_{\bf k}/E_{\bf k})},\nonumber
\end{eqnarray}
where $\xi_{\bf k}=\epsilon_{\bf k}- \mu$, and $\mu$ is chemical potential
 fixed by equation
\begin{equation}
\label{del}
\delta = 2 \sum_{\bf k}
\biggl(v_{\bf k}^2+{{u_{\bf k}^2-v_{\bf k}^2}\over{e^{E_{\bf k}/T}+1}}\biggr)
\end{equation}
All details of approximate analytical and exact numerical solutions
of equation (\ref{BCS}) are discussed in our papers\cite{Flam94,Bel95}.
This equation has an infinite set of solutions of different symmetries
with strongest pairing in the d-wave (four nodes of the gap at the face of
magnetic Brillouin zone). For further discussion we need only
the results for d-wave.
All numerical values are calculated at
$t=3$, $\beta_1/\beta_2=7$, and $f=1.8$. In Table I we present for
different values of hole concentration
1)Fermi energy $\epsilon_F$,
2)chemical potential at zero temperature $\mu(0)$,
3)maximum value of the gap in Brillouin zone at zero temperature
$\Delta_{max}(0)$,
4)maximum value of the gap at Fermi surface at zero temperature
$\Delta_{Fmax}(0)$,
5)chemical potential at critical temperature $\mu(T_c)$,
6)Critical temperature $T_c$.
In Fig.1 we give the map of a single hole mean occupation number
$v_{\bf k}^2$ at $\delta=0.1$ and zero temperature.

\section{Modified spin-wave theory and quantum melting of long-range
antiferromagnetic order}

It is well known that in the long wave-length limit Heisenberg model is
equivalent to the nonlinear $\sigma$-model (e.g., see review
paper\cite{Manousakis}). Therefore usual field theory crossing symmetry
is valid. The same is valid for $t-J$ model. Technically it is evident from
Eq.(\ref{hsw}): at $q \ll 1$ the vertex $g_{\bf k,q} \approx g_{\bf k-q,q}$.
This means that instead of considering a set of Green's functions
$\langle \alpha_{\bf k} \alpha_{\bf k}^{\dag}\rangle$,
$\langle \alpha_{\bf k} \beta_{\bf -k}\rangle$...,
one can introduce one combined Green's function of vector excitation
\begin{equation}
\label{GF}
D(\omega,{\bf q})={{2\omega_{\bf q}}\over{\omega^2-\omega_q^2-
2\omega_{\bf q}\Pi(\omega,{\bf q})}},
\end{equation}
where $\Pi(\omega,{\bf q})$ is mobile holes polarization operator.
For stability of the system the condition
\begin{equation}
\label{stab}
\omega_q + 2 \Pi(0,{\bf q}) > 0
\end{equation}
should be fulfilled. Otherwise the Green's function (\ref{GF}) would
possess poles with imaginary $\omega$. It was demonstrated in the
papers\cite{Shr9,Dom0,Sin0,Ede1,Iga2,Sus3}) that with
$\Pi(0,{\bf q})$ calculated in the normal hole Fermi liquid approximation
the condition (\ref{stab}) is violated. It means the instability of
long-range AF order. Now we want to: 1)Take into account strong hole-hole
pairing, 2)Formulate the approach for description of state without
long-range AF order. In the present section we discuss only point 2).

For elucidation of physical meaning of our approach it is convenient
to use Hamiltonian technique instead of the Feynman one. Due to the
interaction with mobile holes the wave function of the renormalized spin-wave
corresponding to the Green's function (\ref{GF}) is a combination of
$\alpha_{\bf q}^{\dag}$ and $\beta_{\bf -q}$. To find this wave function let us
write down the effective spin-wave Hamiltonian.
\begin{equation}
\label{sw}
H_{sw}=\sum_{\bf q}\biggl((\omega_{\bf q}+\Pi(\omega,{\bf q}))
(\alpha_{\bf q}^{\dag}\alpha_{\bf q}+\beta_{\bf q}^{\dag}\beta_{\bf q})
-\Pi(\omega,{\bf q})
(\alpha_{\bf q}\beta_{\bf -q}+\alpha_{\bf q}^{\dag}\beta_{\bf -q}^{\dag})
\biggr).
\end{equation}
The term proportional to $\omega_{\bf q}$ comes from the ``bare'' Hamiltonian
(\ref{Heff}). First ``$\Pi$-term'' comes from the diagram Fig.3a where one
spin-wave is annihilated and the other is created.
Second ``$\Pi$-term'' comes from diagrams Fig.3bc where
two spin-waves are annihilated or created. Let us note that spin-waves
have definite values of $S_z$: $\alpha_{\bf q}^{\dag}$ has $S_z=-1$
and $\beta_{\bf -q}$ has $S_z=+1$. Therefore they can appear only in
combinations presented in (\ref{sw}). One can certainly prove this
explicitly using (\ref{hsw}) and calculating the polarization
operator. In the second ``$\Pi$-term'' the spin-waves have the opposite
momenta.
At $q \ll 1$ the vertex $g_{\bf k,q}$ is proportional to the momentum $q$.
Just for this reason second ``$\Pi$-term'' has different sign in
comparison with first one.
Diagonalization of the Hamiltonian (\ref{sw}) by the Bogoliubov transformation
gives the spectrum of Bose excitations in the system
\begin{equation}
\label{om}
\Omega_{\bf q}^2=
\omega_{\bf q}^2+2\omega_{\bf q}\Pi(\Omega_{\bf q},{\bf q}).
\end{equation}
This  is exactly the equation for the poles of the Green's function
(\ref{GF}). So the Hamiltonian approach reproduces conventional
result of the Feynman technique.
 Account of $\omega$-dependence of the polarization
operator in Hamiltonian technique is rather cumbersome. Therefore in the
present section we neglect it:
$\Pi(\Omega_{\bf q},{\bf q}) \approx \Pi(0,{\bf q})$.
As usual Bogoliubov transformation gives the ground state of the form
\begin{equation}
\label{gs}
|gs\rangle \propto \exp \biggl(\sum_{\bf q}c_{\bf q}
\alpha_{\bf q}^{\dag}\beta_{\bf -q}^{\dag}\biggr)|0 \rangle,
\end{equation}
with some coefficients $c_{\bf q}$. This is the condensate of spin-waves.

  We started from Neel ground state $|0\rangle$ with two sublattices
$u$-up and $d$-down. The difference in magnetization of the two
sublattices is of the form (see e.g. Ref.\cite{Taka})
\begin{equation}
\label{sz}
{1\over 2}(S_u^z-S_d^z)=0.303-2\sum_{\bf q}{1\over{\omega_{\bf q}}}
\biggl(\alpha_{\bf q}^{\dag}\alpha_{\bf q}+\beta_{\bf q}^{\dag}\beta_{\bf q}
 -\gamma_{\bf q}
(\alpha_{\bf q}\beta_{\bf -q}+\alpha_{\bf q}^{\dag}\beta_{\bf -q}^{\dag})
\biggr).
\end{equation}
Using parameters of transformation
which diagonalizes the Hamiltonian (\ref{sw}) one can easily calculate the
renormalized magnetization
\begin{equation}
\label{magn}
\delta S_z=\langle gs|{1\over 2}(S_u^z-S_d^z)|gs\rangle=0.303-2\int
\biggl({1\over{\Omega_{\bf q}}}-{1\over{\omega_{\bf q}}}\biggr)
{{d^2{\bf q}}\over{(2\pi)^2}},
\end{equation}
where $\Omega_{\bf q}=
\sqrt{\omega_{\bf q}^2+2 \omega_{\bf q} \Pi(0,{\bf q})}$.

In the above discussion we assumed that condition (\ref{stab}) is
fulfilled and hence $\Omega_{\bf q}$ is real. Let us now increase
polarization operator ($\Pi(0,{\bf q}) \to x \cdot \Pi(0,{\bf q})$, $x > 1$)
approaching $\Omega_{\bf q}$ to zero. There are two
possibilities: 1) $\Omega_{\bf q}$ vanishes at some values of ${\bf q}$,
but $\delta S_z$ remains finite positive because of the convergence of the
integral in (\ref{magn}). 2) In approaching $\Omega_{\bf q}$ to zero
$\delta S_z$ vanishes and then becomes negative.
The choice between these two scenario depends on ${\bf q}$-dependence
of $\Pi(0,{\bf q})$. We will demonstrate below that for
paired hole Fermi liquid at $\delta > \delta_c$ second scenario is realized.
Therefore let us discuss this situation.

Vanishing of the magnetization $\delta S_z$ means that there is a
lot of spin-waves in the condensate (\ref{gs}) and we have to take into
account their nonlinear interaction. We can not do it exactly.
Fortunately, there is approximate way.
We can apply the modified spin-wave theory suggested by
Takahashi for the Heisenberg model at nonzero temperature\cite{Taka}.
Following Takahashi let us impose the condition that the sublattice
magnetization vanishes in the quantum spin-liquid state
\begin{equation}
\label{s0}
\langle gs|{1\over 2}(S_u^z-S_d^z)|gs\rangle=0.
\end{equation}
To find the ground state with this condition we have to diagonalize
\begin{equation}
\label{Hnu}
H_{\nu}=H_{sw}-{1\over8}\nu^2(S_u^z-S_d^z),
\end{equation}
where ${1\over8}\nu^2$ is Lagrange multiplier,
$H_{sw}$ is given by (\ref{sw}) and $(S_u^z-S_d^z)$ by (\ref{sz}).
Simple calculation shows that instead of (\ref{om}) we get a spectrum
of excitations with a gap
\begin{equation}
\label{gap}
\Omega_{\nu{\bf q}}=\sqrt{\Omega_{\bf q}^2+\nu^2}.
\end{equation}
The average value of the magnetization is given by the formula (\ref{magn})
with $\Omega_{{\bf q}}$ replaced by $\Omega_{\nu{\bf q}}$. The gap
$\nu$ should be found after substitution of this formula into condition
(\ref{s0}).  Let us stress that this condition reflects strong
nonlinearity of the spin-wave theory. In essence it gives an effective
cutoff of unphysical states in the Dyson-Maleev approach. This
question is discussed in the paper\cite{Dot} where the modified spin-wave
theory  is applied to the description of the transition from AF
state to spin-liquid state in $J_1-J_2$ model.

In the conclusion of present section we would like to note that
realization of the first scenario (finite positive
magnetization $\delta S_z$ at vanishing of $\Omega_{\bf q}$ at some
values of ${\bf q}$) would mean instability with respect to decay to
spiral state. It happens at $\delta < \delta_c$.

\section{Spin-wave Green's functions and polarization operators}

We are interested in hole concentration $\delta \sim 0.15$.
Corresponding value of $q$ is $q \sim p_F \sim 1$.
Therefore, for practical calculations $\sigma$-model long-wave-length
approximation (\ref{GF}) is not enough. The technique which
overcomes this problem was developed by Igarashi and Fulde\cite{Iga22},
see also the paper by Khaliullin and Horsch\cite{Khal3}.
Following these works let us introduce the Green's functions
for spin-waves
\begin{eqnarray}
\label{GFs}
D_{\alpha \alpha}(t,{\bf q})&=&
-i\langle T[\alpha_{\bf q}(t)\alpha_{\bf q}^{\dag}(0)]\rangle, \\
D_{\alpha \beta}(t,{\bf q})&=&
-i\langle T[\alpha_{\bf q}(t)\beta_{\bf -q}(0)]\rangle, \nonumber \\
D_{\beta \alpha}(t,{\bf q})&=&
-i\langle T[\beta_{\bf -q}^{\dag}(t)\alpha_{\bf q}^{\dag}(0)]\rangle,
\nonumber \\
D_{\beta \beta}(t,{\bf q})&=&
-i\langle T[\beta_{\bf -q}^{\dag}(t)\beta_{\bf -q}(0)]\rangle,
\nonumber
\end{eqnarray}
where $T$ is the time-ordering operator, and $\langle . . . \rangle$
represents an average over the exact ground state. In zero approximation
in the interaction with mobile holes the spin-wave Green's functions
in $\omega$-${\bf q}$ representation are
given by\cite{Iga22}
\begin{eqnarray}
\label{GFs0}
D_{\alpha \alpha}^0(\omega,{\bf q})&=&(\omega-\omega_{\bf q}+i0)^{-1},\\
D_{\alpha \beta}^0(\omega,{\bf q})&=&
D_{\beta \alpha}^0(\omega,{\bf q})=0,\nonumber\\
D_{\beta \beta}^0(\omega,{\bf q})&=&(-\omega-\omega_{\bf q}+i0)^{-1}.
\nonumber
\end{eqnarray}
Polarization operator in this technique is also a matrix. So there are
$P_{\alpha \alpha}(\omega,{\bf q})$,
$P_{\alpha \beta}(\omega,{\bf q})$,
$P_{\beta \alpha}(\omega,{\bf q})$, and
$P_{\beta \beta}(\omega,{\bf q})$ components.

In one loop approximation polarization operator is
given by the diagrams presented in Fig.4. where the normal as well as
the anomalous hole Green's function is taken into account. First of all
let us discuss contributions of the ``coherent'' and ``incoherent''
parts of the hole Green's function into the polarization operator.
The diagrams in Fig.4 are represented in terms of the dressed
quasiholes $h_{\bf k}$ which involve in the effective Hamiltonian
(\ref{Heff}). However one can ask the question: what is the accuracy
of this approximation and what happens if we substitute the
Green's function of bare hole $d_{n\sigma}$ into polarization operator?
The Green's function of bare hole is of the form
\begin{equation}
\label{hGF}
G(\epsilon,{\bf k})={{Z}\over{\epsilon - \epsilon_{\bf k}+i0}} + G_{incoh}.
\end{equation}
where $Z$ is quasiparticle residue. If we use bare hole Green's
function we also have to use bare hole-spin-wave coupling constant
$f_{bare}=2t$ in the vertex $g_{\bf k,q}$ (\ref{hsw}). Substitution
of the pole (coherent) part of (\ref{hGF}) into polarization operator
gives combination $f_{bare}Z$, but this is exactly the effective
coupling constant $f$\cite{Suhf}. Thus, the effective theory with
Hamiltonian (\ref{Heff}) is equivalent to the account of the coherent (pole)
part of the bare hole Green's function (\ref{hGF}). According to condition
(\ref{stab}) the most important
characteristic is the polarization operator at zero frequency.
If we neglect the pairing of quasiholes we can easily calculate
its value in normal Fermi-liquid approximation\cite{Sus3})
\begin{equation}
\label{P0q}
P_{\alpha \alpha}(0,{\bf q})=
\Pi(0,{\bf q})\approx -{{\sqrt{2}f^2}\over{\pi\sqrt{\beta_1 \beta_2}}}q,
\ \ \ at \ \ \ q \ll p_F.
\end{equation}
This is coherent contribution. It is independent of the hole concentration!
Incoherent part of the polarization operator which comes from $G_{incoh}$
in (\ref{hGF})is proportional to hole concentration $\delta$. Therefore it is
negligible at $\delta \ll 1$. This conclusion agrees with that of the
Ref.\cite{Beck93}. It is interesting that incoherent contribution
can be also calculated analytically\cite{Mur}
\begin{equation}
\label{Pincoh}
P_{incoh}(0,{\bf q}) \approx -{{f^4}\over{4\sqrt{2}\pi}} \delta
\ln{{1}\over{\delta}} \cdot q.
\end{equation}
At any reasonable $\delta$ the incoherent part $P_{incoh}(0,{\bf q})$ does
not exceed 10\% of $P_{coh}(0,{\bf q})$. Therefore further we neglect
$P_{incoh}$ and use effective theory with the Hamiltonian (\ref{Heff})
which is equivalent to the account of only coherent part in the hole
Green's function (\ref{hGF}).

Due to the symmetry of Green's function it is convenient to introduce
following notations
\begin{eqnarray}
\label{pol}
&&P_{\beta \beta}(-\omega,{\bf q})=
P_{\alpha \alpha}(\omega,{\bf q})=
\Pi(\omega,{\bf q}),\\
&&P_{\alpha \beta}(\omega,{\bf q})=
P_{\beta \alpha}(\omega,{\bf q})={\overline \Pi}(\omega,{\bf q}).\nonumber
\end{eqnarray}
Calculation of the polarization operator Fig.4 in single loop approximation
is straightforward. Account of both normal and anomalous hole Green's
functions gives
\begin{eqnarray}
\label{polpi}
\Pi^{(1)}(\omega,{\bf q})=&&\sum_{\bf k}
\biggl[
v_{\bf k}^2 u_{\bf k+q}^2 \biggl(
{{g^2_{\bf k,q}}\over{\omega-E_{\bf k}-E_{\bf k+q}}}+
{{g^2_{\bf k+q,-q}}\over{-\omega-E_{\bf k}-E_{\bf k+q}}}\biggr)-\\
&&-u_{\bf k} v_{\bf k} u_{\bf -k-q} v_{\bf -k-q}
g_{\bf k,q}g_{\bf k+q,-q}\biggl({1\over{\omega-E_{\bf k}-E_{\bf k+q}}}+
{1\over{-\omega-E_{\bf k}-E_{\bf k+q}}}\biggr)\biggr],\nonumber\\
{\overline \Pi}^{(1)}(\omega,{\bf q})=
&&\sum_{\bf k}
\biggl(v_{\bf k}^2 u_{\bf k+q}^2 g_{\bf k,q}g_{\bf k+q,-q}-
u_{\bf k} v_{\bf k} u_{\bf -k-q} v_{\bf -k-q} g^2_{\bf k,q}\biggr)
\times \nonumber\\
&&\times \biggl({1\over{\omega-E_{\bf k}-E_{\bf k+q}}}+
{1\over{-\omega-E_{\bf k}-E_{\bf k+q}}}\biggr).\nonumber
\end{eqnarray}
In these formulas $g_{\bf k,q}$ is hole-spin-wave vertex (\ref{hsw}), and
$u_{\bf k}$, $v_{\bf k}$, $E_{\bf k}$ are parameters of
BCS hole wave function (\ref{BCS}).

Let us consider now two and more loop corrections to polarization
operator. Some diagrams of this type with
normal hole Green's function are presented in Fig.5. One can easily
prove that the diagram Fig.5a (single spin-wave exchange)  vanishes
due to the spin-flip nature of the hole-spin-wave vertex. Vanishing of
this diagram is absolutely similar to the vanishing of the single loop
correction to the hole-spin-wave vertex (see Refs.\cite{Mart1,Liu2,Suhf}).
The diagram Fig.5b equals to zero for the same reason. To calculate
the diagram Fig.5c let us represent it using Goldstone diagram technique:
Fig.6. In the intermediate state in the diagram Fig.6a as well as in
the diagram Fig.6b there is two-particles two-holes excitation\cite{hole}.
Nevertheless these two diagrams are essentially different.
All intermediate momenta in the diagram Fig.6b are in the vicinity of
Fermi surface (${\bf p= k-k_0} \sim p_F$), and therefore this diagram
is convergent.
On other hand the particles momenta in diagram Fig.6a are not
restricted and this diagram is ultraviolet divergent. The
divergence gives logarithmic enhancement of this diagram.
Further we neglect contribution Fig.6b, and transform
Fig.6a into an effective diagram Fig.7 with ``dot'' given by Fig.8.

We are interested in relatively small $q$ ($q < 1$). Therefore all momenta
in the diagram Fig.7 are close to the Fermi surface. It means that we need
to calculate the effective interaction (``dot'') at Fig.8 at
${\bf k} \approx {\bf k+q} \approx  (\pm \pi/2,\pm \pi/2)$ and
${\bf k^{\prime}} \approx {\bf k^{\prime}+q} \approx (\pm \pi/2,\pm \pi/2)$.
Let us neglect q-dependence (set $q=0$) and denote this effective interaction
by $V^{(1)}({\bf k},{\bf k^{\prime}})$. Single particle state near
${\bf k}_{0--}=(-\pi/2,-\pi/2)$ is related to that near
${\bf k}_{0++}=(\pi/2,\pi/2)$ via umklapp process and therefore these points
are actually equivalent. It is like one pocket of the Fermi surface.
Due to this reason $V^{(1)}({\bf k}_{0++},{\bf k}_{0++})=
V^{(1)}({\bf k}_{0++},{\bf k}_{0--})$. Certainly one can prove this
relation explicitly using perturbation theory expression corresponding
to diagram Fig.8. So there are only two independent pockets of the Fermi
surface (for example centred near ${\bf k}_{0++}$ and ${\bf k}_{0+-}$),
and the effective interaction can be characterized by the two values
only:
\begin{eqnarray}
\label{pok}
&&V^{(1)}_{++}=V^{(1)}({\bf k}_{0++},{\bf k}_{0++}),\\
&&V^{(1)}_{+-}=V^{(1)}({\bf k}_{0++},{\bf k}_{0+-}).\nonumber
\end{eqnarray}
First one is the effective particle-hole interaction when
particle and hole are from the same pocket. Second value
is particle-hole interaction when particle and hole are from different
pockets.

Using hole-spin-wave interaction (\ref{hsw}) one can easily derive
expression for $V^{(1)}_{++}$ corresponding
to  diagram Fig.8
\begin{equation}
\label{dot}
V^{(1)}_{++}=-32f^4\sum_{\bf Q}
{{\gamma^2_{\bf Q+k}\gamma^2_{\bf Q-k}(4-\omega^2_{\bf Q})}\over
{\omega^2_{\bf Q}(\omega_{\bf Q}+E_{\bf Q+k})^2E_{\bf Q+k}}}.
\end{equation}
Here ${\bf k}={\bf k}_{0++}$. At $p_F < Q < \pi$ the integral
in (\ref{dot}) is logarithmic divergent ($\int dQ/Q$) and therefore
$V^{(1)}_{++}$ can be analytically calculated with logarithmic
accuracy
\begin{equation}
\label{dotan}
V^{(1)}_{++} \approx -{{2f^4}\over{\pi \beta_1}}
{{(1+2\sqrt{\beta_2/\beta_1)}}\over{(1+\sqrt{\beta_2/\beta_1})^2}} \cdot L.
\end{equation}
The big logarithm $L$ equals to
\begin{equation}
\label{L}
L=\ln(\epsilon_{\Lambda}/\Delta),
\end{equation}
where $\Delta$ is  typical value of superconducting gap and
$\epsilon_{\Lambda} \sim 2$ is ultraviolet cutoff which is of the
order of hole band width. Actually only this logarithm
justifies introduction of the effective point-like interaction $V^{(1)}$.
Certainly expression (\ref{dot}) can be easily exactly evaluated  numerically.
Similar consideration shows that due to the structure of hole-spin-wave
vertex (\ref{hsw}) effective particle hole interaction for
particle and hole from different pockets vanishes
\begin{equation}
\label{dot+-}
V^{(1)}_{+-}=0
\end{equation}

Above discussion is not the end of the story about effective particle-hole
interaction in the spin-flip channel. It was just a leading order.
To find correct value we have to sum all the ``ladder'' given at Fig.9.
Fortunately each additional rung in ladder gives additional big
logarithm and therefore summation can be easily done with logarithmic
accuracy
\begin{equation}
\label{ladder}
V_{++}=V^{(1)}_{++}\biggl(1+{{f^2}\over{\pi \beta_1
(1+\sqrt{\beta_2/\beta_1})}}\cdot L \biggr)^{-1}.
\end{equation}
Summation of the ladder does not change the relation (\ref{dot+-}):
effective particle hole interaction for particle and hole from different
pockets vanishes.

For very small hole concentration $\delta$ the superconducting gap $\Delta$
is approaching zero, $L \to \infty$, and hence
\begin{equation}
\label{limV}
V_{++} \to -2f^2 {{1+2\sqrt{\beta_2/\beta_1}}\over
{1+\sqrt{\beta_2/\beta_1}}}.
\end{equation}
In practice $L$ is never big enough to reach this limit. Realistic value
of $L$ can be found by comparing (\ref{dotan}) with the result of
exact numerical calculation of (\ref{dot}). At $\delta=0.05$ $L \approx 4$,
and it drops down to $L \approx 3$ at at $\delta=0.2$.
Substituting correct value of $L$ into (\ref{ladder}) we find renormalized
particle-hole interaction in spin-flip channel $V_{++}$. Numerical
values of $V_{++}$ are presented in the Table.II.
Actually it is almost constant $V_{++} \approx -4.5$ for interesting
concentrations.

Now we can calculate exact spin-wave polarization operator chaining
effective particle hole interaction in the spin-flip channel, see Fig.10.
This chaining is substantially simplified by the fact that the pockets
of the Fermi surface are well separated. Let us represent single loop
polarization operator (\ref{polpi}) as a sum of contributions
corresponding to different pockets
\begin{eqnarray}
\label{polpok}
&&\Pi^{(1)}=\Pi^{(1)}_I+\Pi^{(1)}_{II},\\
&&\Pi^{(1)}_I(\omega,{\bf q})=\sum_{{\bf k}\in I}X(\omega,{\bf k},{\bf q}),
\nonumber\\
&&\Pi^{(1)}_{II}(\omega,{\bf q})=\sum_{{\bf k}\in II}X(\omega,{\bf k},{\bf q}),
\nonumber
\end{eqnarray}
where $X(\omega,{\bf k},{\bf q})$ is the integrand in formula (\ref{polpi}).
Regions $I$ and $II$ of magnetic Brillouin zone are shown at Fig.1. We
remind that point ${\bf k}_{0--}=(-\pi/2,-\pi/2)$ is connected with
${\bf k}_{0++}=(\pi/2,\pi/2)$ via umklapp process and therefore
the region $I$ represents one pocket. The same is valid for $II$.
Due to the condition (\ref{dot+-}) particle-hole scattering does not
transfer particle-hole excitation from one pocket to another, but
inside pocket scattering amplitude (\ref{ladder}) is constant. In this
situation summation of chain in Fig.10 is trivial. The exact spin-wave
polarization operator (\ref{pol}) is of the form
\begin{eqnarray}
\label{polex}
&&\Pi(\omega,{\bf q})={{\Pi^{(1)}_I(\omega,{\bf q})}\over
{1+V_{++}Q_I(\omega,{\bf q})}}+
{{\Pi^{(1)}_{II}(\omega,{\bf q})}\over
{1+V_{++}Q_{II}(\omega,{\bf q})}},\\
&&{\overline \Pi}(\omega,{\bf q})={{{\overline \Pi}^{(1)}_I(\omega,{\bf q})}
\over{1+V_{++}Q_I(\omega,{\bf q})}}+
{{{\overline \Pi}^{(1)}_{II}(\omega,{\bf q})}\over
{1+V_{++}Q_{II}(\omega,{\bf q})}},\nonumber
\end{eqnarray}
where
\begin{equation}
\label{Q}
Q_{I,II}(\omega,{\bf q})=\sum_{{\bf k}\in I,II}
\biggl(v_{\bf k}^2 u_{\bf k+q}^2+
u_{\bf k} v_{\bf k} u_{\bf -k-q} v_{\bf -k-q}\biggr)
 \biggl({1\over{\omega-E_{\bf k}-E_{\bf k+q}}}+
{1\over{-\omega-E_{\bf k}-E_{\bf k+q}}}\biggr).
\end{equation}

Unfortunately further analytical calculation of $\Pi(\omega,{\bf q})$ and
${\overline \Pi}(\omega,{\bf q})$ given by (\ref{polex}) is
impossible. However numerical calculation using (\ref{polpi}),
(\ref{polpok}),(\ref{Q}) and parameters of superconducting
wave function found in the Section II is straightforward.
We will discuss it in detail in the next section. Let us look
here only at the limit $q \to 0$. In this limit
\begin{equation}
\label{q=0}
\Pi(\omega,{\bf q})=\Pi(-\omega,{\bf q})=-{\overline \Pi}(\omega,{\bf q}).
\end{equation}
This reflects the fact that we can forget about all
complications connected with multi component spin-wave Green's function
(\ref{GFs}), and use simple $\sigma$-model picture (\ref{GF}).
Let us introduce parameter of spin-wave instability
\begin{equation}
\label{zakr}
C= {{-2\Pi(0,{\bf q})}\over{\omega_{\bf q}}}, \ {\bf q}\to 0.
\end{equation}
According to equation (\ref{stab}) if $C > 1$ spectrum of
excitations is unstable. Naive value of $C$ corresponding to normal
Fermi liquid approximation (\ref{P0q}) is
\begin{equation}
\label{naive}
C_{naive}={{2f^2}\over{\pi \sqrt{\beta_1 \beta_2}}} \approx 2.8.
\end{equation}
Numerical values of $C$ obtained with (\ref{polex}) for different hole
concentrations are presented in the Table II. We see that pairing
and particle-hole rescattering (\ref{ladder}) substantially
reduce value of $C$. Actually rescattering is more important, but
it has the same origin as the pairing. Nevertheless $C>1$, and long-range
antiferromagnetic order is unstable.

\section{Spin-wave Green's function.}

To deal with spin-wave instability let us apply approach developed
in the Section III. With Takahashi condition (\ref{s0}),(\ref{sz})
bare spin-wave Hamiltonian is transformed to
\begin{eqnarray}
\label{Hnu1}
&&\sum_{\bf q}\omega_{\bf q}(\alpha_{\bf q}^{\dag}\alpha_{\bf q}
+\beta_{\bf q}^{\dag}\beta_{\bf q}) \to
\sum_{\bf q}\omega_{\bf q}(\alpha_{\bf q}^{\dag}\alpha_{\bf q}
+\beta_{\bf q}^{\dag}\beta_{\bf q}) -
{1\over8}\nu^2(S_u^z-S_d^z) \to \nonumber\\
&&\to \sum_{\bf q}\biggl[\biggl(\omega_{\bf q}+{{\nu^2}\over{2\omega_{\bf q}}}
\biggr)(\alpha_{\bf q}^{\dag}\alpha_{\bf q}+\beta_{\bf q}^{\dag}\beta_{\bf q})
-\gamma_{\bf q}{{\nu^2}\over{2\omega_{\bf q}}}
(\alpha_{\bf q}\beta_{\bf -q}+\alpha_{\bf q}^{\dag}\beta_{\bf -q}^{\dag})
\biggr].
\end{eqnarray}
Lagrange multiplier $\nu$ will be found later from the
condition (\ref{s0}). Dioganalization of the Hamiltonian (\ref{Hnu1})
by usual Bogoliubov transformation gives new bare spin-wave
spectrum
\begin{equation}
\label{omnu}
\omega_{\nu {\bf q}}^2=\omega_{\bf q}^2+\nu^2+{{\nu^4}\over{16}}
\approx \omega_{\bf q}^2+\nu^2,
\end{equation}
and new bare spin-wave operators $A_{\bf q}^{\dag}$, $B_{\bf q}^{\dag}$
\begin{eqnarray}
\label{AB}
&&\alpha_{\bf q}=c_{\bf q}A_{\bf q} + s_{\bf q}B_{\bf -q}^{\dag},\\
&&\beta_{\bf -q}=s_{\bf q}A_{\bf q}^{\dag} + c_{\bf q}B_{\bf -q},\nonumber\\
&&c_{\bf q},s_{\bf q}
={{\omega_{\nu {\bf q}} \pm \omega_{\bf q}}\over
{2 \omega_{\nu {\bf q}} \omega_{\bf q}}},\nonumber\\
\end{eqnarray}
We will see below that $\nu$ is very small. Therefore we neglect
$\nu^4/16$ in comparison with $\nu^2$. Hole-spin-wave interaction
(\ref{hsw}) should be also expressed in terms of $A_{\bf q}$ and
$B_{\bf q}$ operators
\begin{eqnarray}
\label{HSW}
&& H_{h,sw} = \sum_{\bf k,q} G_{\bf k,q}
    \biggl(
  h_{{\bf k}+{\bf q}\downarrow}^{\dag} h_{{\bf k}\uparrow} A_{\bf q}
  + h_{{\bf k}+{\bf q}\uparrow}^{\dag} h_{{\bf k}\downarrow} B_{\bf q}
  + \mbox{H.c.}   \biggr),\\
&& G_{\bf k,q} = c_{\bf q}g_{\bf k,q}+s_{\bf q}g_{\bf k+q,-q}.\nonumber
\end{eqnarray}
Let us note that due to the Goldstone theorem $g_{\bf k,q} \to 0$ at
${\bf q} \to 0$. However, in the state which we discuss now rotational
symmetry is restored and therefore $G_{\bf k,0}=\sqrt{2}f\gamma_{\bf k}$.

Similar to (\ref{GFs}) we introduce Green's functions $D_{AA}(t,{\bf q})$,
$D_{AB}(t,{\bf q})$, $D_{BA}(t,{\bf q})$ and $D_{BB}(t,{\bf q})$.
Zero approximation Green's functions $D_{AA}^0(\omega,{\bf q})$...
are given by eqs. (\ref{GFs0}) with replacement $\omega_{\bf q} \to
\omega_{\nu{\bf q}}$. Polarization operators are given by the formulas
of the previous section with replacement $g_{\bf k,q} \to G_{\bf k,q}$.
Spin-wave Green's functions obey usual Dyson equations\cite{Iga22,Khal3})
\begin{eqnarray}
\label{Dyson}
&&D_{AA}=D^{(0)}_{AA}+
D^{(0)}_{AA}P_{AA}D_{AA}+D^{(0)}_{AA}P_{AB}D_{BA},\\
&&D_{BB}=D^{(0)}_{BB}+
D^{(0)}_{BB}P_{BB}D_{BB}+D^{(0)}_{BB}P_{BA}D_{AB},\nonumber\\
&&D_{AB}=
D^{(0)}_{AA}P_{AA}D_{AB}+D^{(0)}_{AA}P_{AB}D_{BB},\nonumber\\
&&D_{BA}=
D^{(0)}_{BB}P_{BA}D_{AA}+D^{(0)}_{BB}P_{BB}D_{BA}.\nonumber
\end{eqnarray}
Solution of these equations together with (\ref{pol}) gives
\begin{eqnarray}
\label{Dren}
&&D_{AA}(\omega,{\bf q})=D_{BB}(-\omega,{\bf q})=
{{-\omega-\omega_{\bf q}-\Pi(-\omega)}\over
{\lambda(\omega,{\bf q})}},\\
&&D_{AB}(\omega,{\bf q})=
D_{BA}(\omega,{\bf q})=
{{{\overline \Pi}(\omega)}\over{\lambda(\omega,{\bf q})}},\nonumber\\
&&\lambda(\omega,{\bf q})=\nonumber\\
&&=-\omega^2+\omega^2_{\nu{\bf q}}+
\omega_{\nu{\bf q}}\biggl(\Pi(\omega)+\Pi(-\omega)\biggr)+
\omega\biggl(\Pi(\omega)-\Pi(-\omega)\biggr)+
\Pi(\omega)\cdot\Pi(-\omega)-
{\overline \Pi}^2(\omega).
\end{eqnarray}
In these formulas we omit for simplicity argument ${\bf q}$ in the
polarization operator. We do not specify imaginary part of
$\lambda$ in the poles because below we perform Wick's rotation
$\omega \to i\xi$ which automatically gives correct behaviour.
Using (\ref{sz}) one can easily express average magnetization in
the terms of Green's functions (\ref{Dren})
\begin{eqnarray}
\label{m1}
0&=&\langle {1\over 2}(S_u^z-S_d^z)\rangle=
1-2i\sum_{\bf q}{1\over{\omega_{\nu{\bf q}}}}
 \biggl\{D_{AA}(t=-0,{\bf q})+D_{BB}(t=-0,{\bf q})-\nonumber\\
&-&\gamma_{\bf q}\biggl(D_{AB}(t=-0,{\bf q})+D_{BA}(t=-0,{\bf q})\biggr)
\biggr\}=\\
&=&1-2\sum_{\bf q}{1\over{\Omega_{\nu{\bf q}}}}
=0.303-2\sum_{\bf q}\biggl({1\over{\Omega_{\nu{\bf q}}}}-
{1\over{\omega_{\bf q}}}\biggr),\nonumber
\end{eqnarray}
where $\Omega_{\nu{\bf q}}$ is given by the expression
\begin{equation}
\label{m2}
{1\over{\Omega_{\nu{\bf q}}}}=
{1\over{\omega_{\nu{\bf q}}}}
\int_{-\infty}^{+\infty}{{d\xi}\over{2\pi}}
{{1}\over{ \lambda(i\xi,{\bf q})}}\biggl(
2\omega_{\nu{\bf q}}+[\Pi(i\xi,{\bf q})+\Pi(-i\xi,{\bf q})
+2\gamma_{\bf q}{\overline \Pi}(i\xi,{\bf q})]\biggr).
\end{equation}
We have to find the Lagrange multiplier $\nu$ from
equation (\ref{m1}) . At small
${\bf q}$, due to the relation (\ref{q=0}), the expression in square
brackets in (\ref{m2}) vanishes and $\lambda(i\xi,{\bf q})
\to \xi^2+\omega^2_{\nu{\bf q}}+2\omega_{\nu{\bf q}}\Pi(i\xi,{\bf q})$.
If one neglects also $\xi$-dependence of polarization operator,
the integration over $\xi$ in (\ref{m2}) is trivial and it is reduced
to formula (\ref{gap}) derived in Section III in $\sigma$-model
approximation.

 Now we can proceed to the discussion of the results of numerical
calculations. Dependence of hole-hole pairing on the modification
of spin-wave spectrum is rather weak at $\delta \ll 1$, and we
neglect it (see also end of these section). Therefore, we use parameters of
hole BCS wave function
found in Section II. Calculation of the polarization operator
is described in detail in Section IV. One should not forget about
the replacement $g_{\bf k,q} \to G_{\bf k,q}$, but numerically it is
not very important. The values of $\nu$ found from equation (\ref{m1})
are presented in the Table II.
The effective spin-wave frequency $\Omega_{\nu{\bf q}}$ depends both
on direction and magnitude of ${\bf q}$. It is convenient to take the
bare spin-wave frequency $\omega_{\bf q}$ as an argument instead of
$|{\bf q}|$. The plots of $\Omega_{\nu{\bf q}}$ as a function of
$\omega_{\bf q}$ for the directions ${\bf q}\propto (1,0)$ and
${\bf q}\propto (1,1)$ and for different hole concentrations
are given in Figs.11a-d.

At very small hole concentration $\delta$, equation (\ref{m1}) has
no solution. One can easily understand the reason for this.
The parameter of spin-wave instability $C$ is larger then 1 at
arbitrary small hole concentration $\delta$, see
eqs.(\ref{zakr}),(\ref{naive}) and Table II. It means that dependence of
$\Omega_{\nu{\bf q}}$ on $|{\bf q}|$ should be of the type of
solid line at Fig.11a with minimum at $q_0 \sim p_F$. Let us denote
the value of $\Omega_{\nu{\bf q}}$ at this point by $\Omega_0$.
This minimum value is directly related to $\nu$. Therefore instead of
$\nu$ we can consider $\Omega_0$ as a variable which should be found
from eq.(\ref{m1}). If frequency $\Omega_{\nu{\bf q}}$ is isotropic
$\Omega^2_{\nu{\bf q}}\approx \Omega^2_0+a(q-q_0)^2$ the equation
(\ref{m1}) has obvious solution at arbitrary small $\delta$
\begin{equation}
\label{Om0}
\Omega_0 \sim e^{-1/q_0}\sim e^{-1/p_F}\sim e^{-1/\sqrt{\delta}}
\end{equation}
The problem is that the polarization operator and hence
$\Omega^2_{\nu{\bf q}}$ is not isotropic. It is seen explicitly
from Fig. 11a. The anisotropy is proportional to some power of
$q_0 \propto \sqrt{\delta}$. Therefore at small enough concentration
the anisotropy is larger than exponentially small $\Omega_0$ and
equation (\ref{m1}) has no solution. Thus if we start from spin-liquid
phase (say from $\delta=0.1$) and decrease hole density $\delta$,  at
some critical value $\delta=\delta_c$ the gap $\Omega_0$ vanishes.
It means instability of spin liquid state with respect to decay to
spin waves with momentum ${\bf q_0}$ (spirals).
The critical value $\delta_c \approx 3-4\%$. Due to the strong
exponential dependence of (\ref{Om0}) on $\delta$ it is hard to find
numerically the exact value of $\delta_c$. We would like to note that
the dip in Fig.11a is a result of virtual admixture of
spirals to the spin-liquid ground state.

One can consider $\nu$ as a gap $\Delta_M$ in the spin-wave spectrum at
$q=0$. At small hole concentration ($\delta=0.05,0.1$) minimum
of the effective spectrum is shifted from the point $q=0$,
see Figs.11ab. One has to be careful with interpretation
of $\Delta_M$. The effective frequency $\Omega^2_{\nu{\bf q}}$
defined by eq.(\ref{m2}) is not a frequency of simple excitation in
the sense that Green's function
$G \propto 1/(\omega^2-\Omega^2_{\nu{\bf q}})$. Green's function is
of essentially more complicated form. It  has poles and cuts.
Therefore, using standard terminology, it is better to say that $\Delta_M$
is pseudo-gap and that $\Omega_{\nu{\bf q}}$ is pseudo-spectrum.
The calculated value of pseudo-gap as well as its dependence on the
hole concentration reasonably agrees with experimental
data\cite{Tran,Ros,Sato,Reg}.

 In equation (\ref{m1}) we have overcome the problem of
real spectrum by integrating
over imaginary $\omega$. For real $\omega$ structure of the
spectrum is as follow.  At very small $q\ll p_F$ there is a
damped spin-wave with frequency $\omega \approx \Delta_M \approx \nu$.
The decay of the spin-wave into particle-hole pair is allowed due
to the nodes of the superconducting gap.
With increasing of $q$, spectrum is split into the two branches:
1)damped spin-wave with frequency $\omega \sim \omega_{\bf q}$,
and 2)low frequency collective excitation consisting mainly of the particles
and holes. Starting from some value of $q$ decay of this excitation
into particle and hole is forbidden due to the superconducting gap.
Therefore in this low frequency region spin-wave Green's function can be
represented as
\begin{equation}
\label{col}
G(\omega,{\bf q})={{Z}\over{\omega^2-o^2_{\bf q}}},
\end{equation}
where $Z$ is the spin-wave residue. The frequency of collective
excitation $o_{\bf q}$ as a function of $\omega_{\bf q}$ for
direction ${\bf q} \propto (1,0)$ and for hole concentration
$\delta=0.1$ is plotted at Fig.11b. Due to the decay to real
particle-hole pair this spectrum has no meaning at very small $q$
and at large $q$. Spin-wave residue $Z$ changes along the drawn
curve from $Z=0.75$ at left hand side of curve to $Z=0.15$ at
right hand side of the interval. Detail analysis of the
spectrum of spin $S=1$ excitations is  an important problem.
It requires calculation of the dynamic magnetic formfactor
$S_M(\omega,{\bf q})$ of the system. Having this formfactor one
can perform accurate comparison with an experimental data
on neutron scattering and NMR.  It is a subject of a separate work.
Here we concentrate on the magnetic correlation length $\xi_M$ and
static magnetic formfactor $S_M({\bf q})$.

This is the place to remind the reader that in the calculation of the
superconducting pairing we have used unperturbed spin-wave spectrum.
This approximation is justified by the parameter $\sqrt{\delta} \ll 1$
(see Refs.\cite{Flam94,Bel95} and eq.(\ref{omega})).
However for most interesting concentrations
($\delta \sim 0.15-0.2$) this parameter is not so good. We see from
Fig.11a that for $\delta=0.05$ the effective spin-wave spectrum is actually
close to the unperturbed one. However for $\delta=0.2$ (Fig.11d)
deviation is rather big. Therefore to complete our program we have to solve
Eliashberg equations for pairing with renormalized spin-wave Green's
function (compare with Ref.\cite{Plak}). This should be a subject
of a separate work. However qualitatively the influence of this effect is
evident. Effective pairing interaction
(\ref{Vkk}) is roughly proportional to the inverse spin-wave frequency.
Therefore, the renormalized interaction  gives stronger pairing than that in
``zero approximation''. Due to our preliminary estimations, the
correction to pairing is about 30\% at $\delta =0.2$.

\section{Magnetic correlation length and
static magnetic formfactor}

We remind the reader that the spin wave operators $\alpha_{\bf q}$
and $\beta_{\bf q}$ are introduced using Dyson-Maleev
transformation\cite{DM} for localized spin ${\bf S}$,
\begin{eqnarray}
\label{dm}
&&S_l^-=a_l^{\dag}, \ \ \ S_l^+=(2S-a_l^{\dag}a_l)a_l,\nonumber\\
&&S_l^z=S-a_l^{\dag}a_l, \ \ \ for\ \ l \in up \ \ \ sublattice;\\
&&S_m^-=b_m, \ \ \ S_m^+=b_m^{\dag}(2S-b_m^{\dag}b_m),\nonumber\\
&&S_m^z=-S+b_m^{\dag}b_m, \ \ \ for\ \ m \in dawn \ \ \ sublattice.
\nonumber
\end{eqnarray}
The operators $\alpha_{\bf q}$ and $\beta_{\bf q}$ are related to
$a_l$ and $b_m$ by usual combination of Fourier and Bogoliubov
transformations, see e.g. Refs.\cite{Taka,Manousakis}. Following
Takahashi\cite{Taka} we introduce notations
\begin{eqnarray}
\label{fg}
f({\bf r}_l-{\bf r}_{l^{\prime}})&=&
\langle a_l^{\dag}a_{l^{\prime}}\rangle,
\ \ \ l \ne l^{\prime}, \nonumber\\
f({\bf r}_m-{\bf r}_{m^{\prime}})&=&
\langle b_m^{\dag}b_{m^{\prime}}\rangle, \ \ \
m \ne m^{\prime},\\
g({\bf r}_l-{\bf r}_m)&=&
\langle a_l^{\dag}b_m^{\dag}\rangle=
\langle a_l b_m\rangle, \nonumber
\end{eqnarray}
where brackets $\langle . . . \rangle$ represents an average over
ground state. Using mean field procedure for the averaging of the
quartic terms ($\langle a^{\dag}a b^{\dag}b\rangle=
\langle a^{\dag}a\rangle \langle b^{\dag}b\rangle+
\langle a^{\dag}b^{\dag}\rangle \langle ab\rangle$) one can
express spin-spin correlators  in terms of functions $f$ and $g$
(see Ref.\cite{Taka}).
\begin{eqnarray}
\label{SS}
\langle {\bf S}_i {\bf S}_j \rangle &=&f^2({\bf r}_i-{\bf r}_j)
-{1\over{4}}\delta_{i,j},\ \ \
i,j \in same\ \ sublattice,\\
\langle {\bf S}_i {\bf S}_j \rangle &=&-g^2({\bf r}_i-{\bf r}_j),\ \ \
i,j \in different\ \ sublattices \nonumber.
\end{eqnarray}

Using definition (\ref{fg}), the functions $f$ and $g$ can be
expressed in terms of Green's functions $D_{AA}(t=-0,{\bf q})$,
$D_{AB}(t=-0,{\bf q})$ ... This is quite similar to the average
magnetization (\ref{m1}). Then with the help of formulas (\ref{Dren})
one gets
\begin{eqnarray}
\label{fgexp}
f({\bf r})&=&
2\sum_{\bf q}{{e^{i{\bf qr}}} \over{\Omega_{\nu{\bf q}}}},\\
g({\bf r})&=&
-2\sum_{\bf q}{{e^{i{\bf qr}}}
\over{\Omega^{\prime}_{\nu{\bf q}}}},\nonumber
\end{eqnarray}
The effective frequency $\Omega_{\nu{\bf q}}$ is given by equation
(\ref{m2}). The ``prime'' effective frequency is defined by
slightly different formula
\begin{equation}
\label{m3}
{1\over{\Omega^{\prime}_{\nu{\bf q}}}}=
{1\over{\omega_{\nu{\bf q}}}}
\int_{-\infty}^{+\infty}{{d\xi}\over{2\pi}}
{{1}\over{ \lambda(i\xi,{\bf q})}}\biggl(
\gamma_{\bf q}[2\omega_{\nu{\bf q}}+\Pi(i\xi,{\bf q})+\Pi(-i\xi,{\bf q})]
+2{\overline \Pi}(i\xi,{\bf q})\biggr).
\end{equation}
For small momentum $\gamma_{\bf q} \approx 1$ . Therefore
$\Omega^{\prime}_{\nu{\bf q}} \approx \Omega_{\nu{\bf q}}$ for $q \ll 1$.
According to computations the relation
$1/\Omega^{\prime}_{\nu{\bf q}} \approx \gamma_{\bf q}/\Omega_{\nu{\bf q}}$
is valid with reasonable accuracy for arbitrary $q$.
Let us note that due to the spin-liquid equation (\ref{m1}) $f(0)=1$.
Together with (\ref{SS}) this gives correct value of spin:
$\langle {\bf S}^2 \rangle =3/4$.

The results of numerical calculation of the functions $f$ and $g$
are given in Fig.12. Distance between sites attains the values
$r=\sqrt{m^2+n^2}$. We plot $\ln{g(r)}$ for odd value of $m+n$,
and $\ln{f(r)}$ for even value of $m+n$. One can see that for
$r > 2$ the dependence is practically linear. Hence
\begin{equation}
\label{SSxi}
\langle {\bf S}(t,{\bf r})\cdot {\bf S}(t,0) \rangle \propto
(-1)^{m+n}\exp(-r/\xi_M), \ \ \ for \ r > 2.
\end{equation}
The values of the magnetic correlation length $\xi_M$ found from
Fig.12 are given in the Table II.

Using eqs.(\ref{SS}) and (\ref{fgexp}) one can express static
magnetic formfactor in terms of the effective frequencies
$\Omega_{\nu{\bf k}}$ and $\Omega^{\prime}_{\nu{\bf k}}$
\begin{equation}
\label{ffact}
S_M({\bf q})=\sum_{\bf r}\exp(i{\bf q}\cdot {\bf r})
\langle {\bf S}(t,{\bf r})\cdot {\bf S}(t,0)=
-{1\over{4}}+2\sum_{\bf k}\biggl(
{1\over{\Omega_{\nu{\bf k}}\Omega_{\nu{\bf k+q}}}}-
{1\over{\Omega^{\prime}_{\nu{\bf k}}\Omega^{\prime}_{\nu{\bf k+q}}}}
\biggr)
\end{equation}
For ${\bf q}=0$ the integrand in (\ref{ffact}) vanishes at small ${\bf k}$
because $\Omega_{\nu{\bf k}} =\Omega^{\prime}_{\nu{\bf k}}$ at $k \ll 1$.
Therefore only large momenta contribute into $S_M(0)$. For
${\bf q}={\bf Q}=(\pi,\pi)$ situation is different because
$\Omega_{\nu{\bf k+Q}} =\Omega_{\nu{\bf k}}$ and
$\Omega^{\prime}_{\nu{\bf k+Q}} =-\Omega^{\prime}_{\nu{\bf k}}$.
Therefore contribution of small $k$ into $S_M({\bf Q})$ is stressed like
one over spin-wave pseudo-gap squared ($1/\nu^2$). The values of $S_M(0)$ and
$S_M({\bf Q})$ for different hole concentrations are given in the Table II.
The static formfactor $S_M({\bf q})$ at hole concentration $\delta=0.15$
is plotted at Fig.13.
We have to note that due to the strong dependence on the pseudo-gap
which is nearly zero at $\delta=0.05$, the value of $S_M({\bf Q})$
at this concentration is  sensitive to  small variation of parameters.
Therefore the accuracy of calculation of $S_M({\bf Q})$ at this point
is very poor. This is reflection of the closeness to the critical
concentration $\delta_c$ at which system drops down into the irregular
spin-glass state (see discussion in the previous Section).

Although we do not observe the shift of the $S_M({\bf q})$ maximum
from the point ${\bf Q}=(\pi,\pi)$ our result agrees qualitatively
with that of finite cluster numerical simulation\cite{Dag0}
(see also review paper\cite{Dag4} and references therein, where
results for other models are presented). However the values of $S_M({\bf Q})$
obtained by us are substantially higher than those from finite cluster
simulations. We have mentioned already that $S_M({\bf Q})$ is very
sensitive to the spin-wave pseudo-gap at small momenta.
Therefore the possible reason of disagreement is that with relatively small
claster it is impossible to reproduce correct spin-wave spectrum at
small momenta.

\section{Conclusion}

In the present work using the picture of spin-wave condensation
we have derived explicitly spin liquid ground state of the doped
two-dimensional $t-J$ model. Translation invariant
spin-liquid solution exists at doping
$\delta \ge \delta_c \approx 3-4\%$. We have demonstrated that spin-wave
condensation and condensation of Cooper pairs (superconducting pairing of
mobile holes) are intrinsically connected. The developed approach
allows to calculate any property of the ground state as well as
spectrum of excitations. In the present work we have calculated
spin-wave pseudo-gap $\Delta_M$ (Table II), effective spectrum
(or pseudo-spectrum) of
$S=1$ excitations (Fig.11), magnetic correlation length $\xi_M$ and
static magnetic formfactor $S_M(\bf q)$ (Table II and Fig. 13).

\section{ACKNOWLEDGMENTS}

We have derived the basic results presented in the section III
together with M. Yu. Kuchiev.
I am also very grateful to him for numerous helpful discussions.
I thank V. V. Flambaum, G. F. Gribakin and A. V. Dotsenko for
critical and stimulating discussions.

\newpage

\begin{table}
\caption{d-wave pairing at $t=3$, $\beta_1/\beta_2=7$, $f=1.8$
and different values of hole concentration $\delta$.
Units correspond to $J=1$.
The values of 1)Fermi energy $\epsilon_F$,
2)Chemical potential at zero temperature $\mu(0)$,
3)Maximum value of the gap in Brillouin zone at zero temperature
$\Delta_{max}(0)$,
4)Maximum value of the gap at Fermi surface at zero temperature
$\Delta_{Fmax}(0)$,
5)Chemical potential at critical temperature $\mu(T_c)$,
6)Critical temperature $T_c$. For guidance we give in brackets
the value in degrees, assuming that $J=0.15$eV. }

\begin{tabular}{l l l l l l l }
\multicolumn{1}{c}{$\delta$} &
\multicolumn{1}{c}{$\epsilon_F$} &
\multicolumn{1}{c}{$\mu(0)$} &
\multicolumn{1}{c}{$\Delta_{max}(0)$} &
\multicolumn{1}{c}{$\Delta_{Fmax}(0)$} &
\multicolumn{1}{c}{$\mu(T_c)$} &
\multicolumn{1}{c}{$T_c$} \\ \hline

0.05& 0.054 & 0.049 & 0.033 & 0.031 & 0.053 & 0.016 (28$^o$) \\
0.1 & 0.103 & 0.091 & 0.067 & 0.056 & 0.097 & 0.029 (50$^o$) \\
0.15& 0.144 & 0.125 & 0.102 & 0.079 & 0.130 & 0.040 (70$^o$) \\
0.20& 0.178 & 0.155 & 0.130 & 0.097 & 0.160 & 0.048 (84$^o$) \\
\end{tabular} \end{table}

\begin{table}
\caption{Parameters relevant to spin-wave spectrum at
$t=3$, $\beta_1/\beta_2=7$, $f=1.8$
and different values of hole concentration $\delta$.
Units correspond to $J=1$.
1) Renormalized particle-hole interaction in spin-flip channel $V_{++}$.
2) The parameter of spin-wave instability $C$ defined by (41).
3) Lagrange parameter $\nu$  which is equal to the
pseudo-gap $\Delta_M$ in the spin-wave spectrum. In the brackets we
give the value in degrees, assuming that $J=0.15$eV.
4) Magnetic correlation length $\xi_M$ in the units of lattice spacing.
5) Static magnetic formfactor $S_M({\bf q})$ for ${\bf q}=0$.
6) Static magnetic formfactor $S_M({\bf q})$ for ${\bf q}={\bf Q}=
(\pi,\pi)$}

\begin{tabular}{l l l l l l l l l}
\multicolumn{1}{c}{$\delta$} &
\multicolumn{1}{c}{$V_{++}$} &
\multicolumn{1}{c}{$C$}      &
\multicolumn{1}{c}{$\nu=\Delta_M$}    &
\multicolumn{1}{c}{$\xi_M$}  &
\multicolumn{1}{c}{$S(0)$}  &
\multicolumn{1}{c}{$S_M({\bf Q})$} \\ \hline

0.05& -5.0 & 1.18 & 0.048 (84$^o$)& 3.4 & 0.035 & $\sim 12.$\\
0.1 & -4.7 & 1.14 & 0.057 (99$^o$)& 2.2 & 0.066 & \ \ \ 7.9\\
0.15& -4.5 & 1.10 & 0.075 (130$^o$)& 1.6 & 0.088 & \ \ \ 5.0\\
0.20& -4.3 & 1.05 & 0.098 (171$^o$)& 1.3 & 0.11  & \ \ \ 3.7\\
\end{tabular} \end{table}

{\bf FIGURE CAPTIONS}

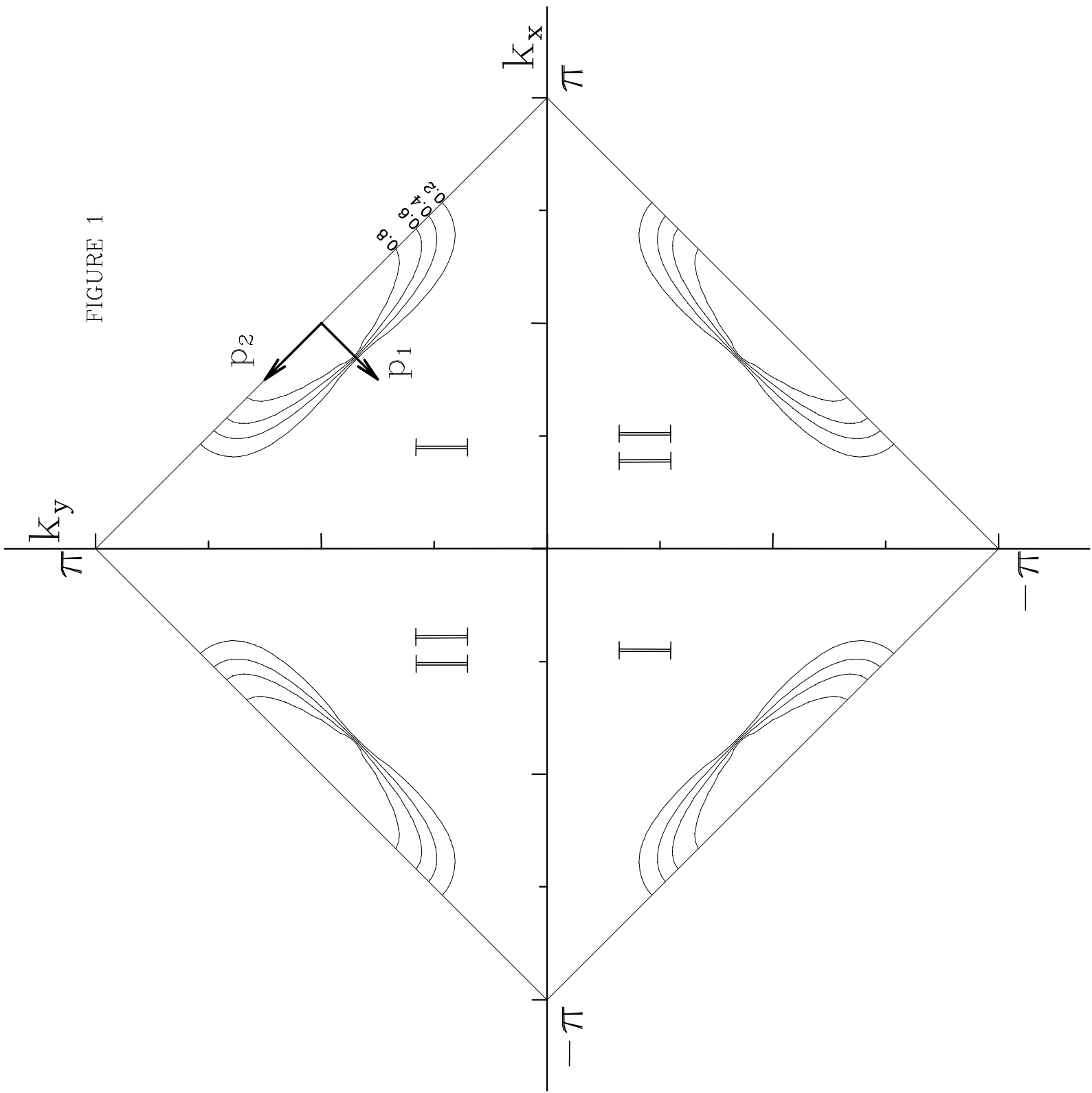
FIG. 1. Magnetic Brillouin zone and the contour plot of a single hole
mean occupation number $v_{\bf k}^2$  at $\delta=0.1$ and zero
temperature.\\
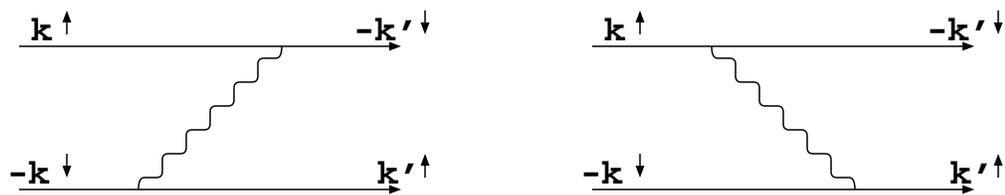
FIG. 2. Single spin-wave exchange mechanism of hole-hole attraction.\\
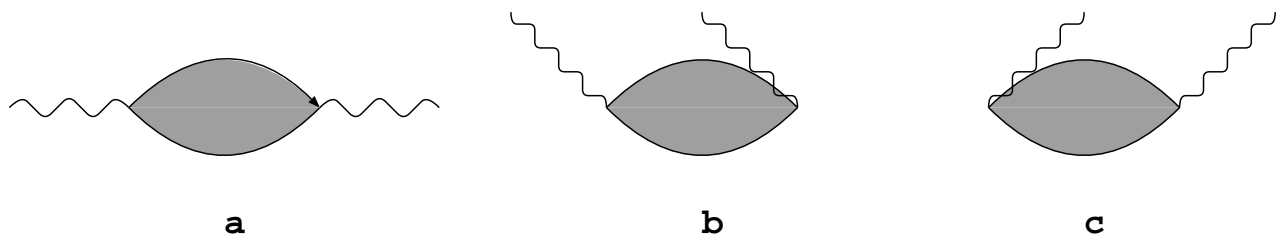
Fig. 3.The spin-wave polarization operator in the Schrodinger
representation:\\
a)One spin-wave is annihilated and the other one is created.\\
b,c)Two spin-waves are either annihilated or created.\\
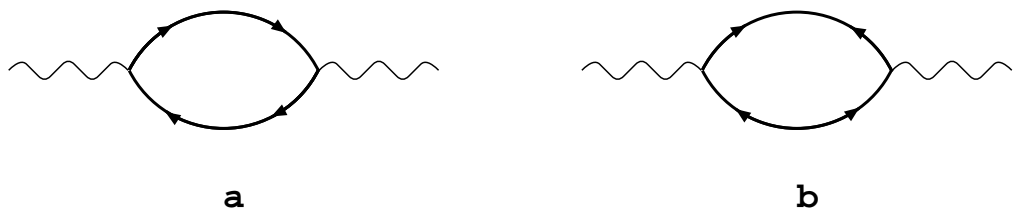
FIG. 4. Polarization operator $P_{\alpha \alpha}(\omega,{\bf q})$ in one
loop approximation. a - contribution of the normal hole Green's function,
b - contribution of the anomalous hole Green's function.\\
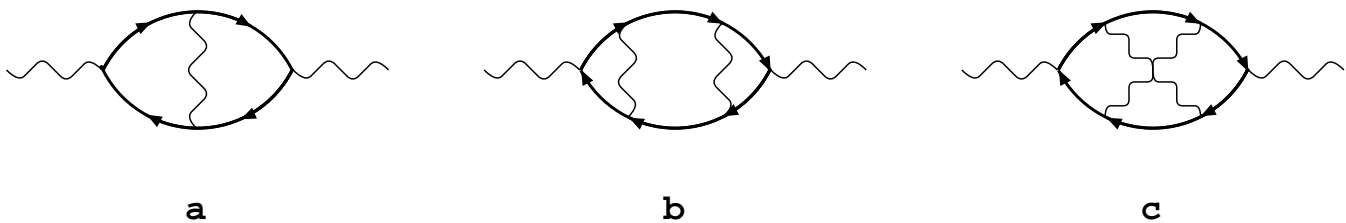
FIG. 5. Corrections to the polarization operator. a)Single spin-wave
exchange, b,c)double spin wave exchange.\\
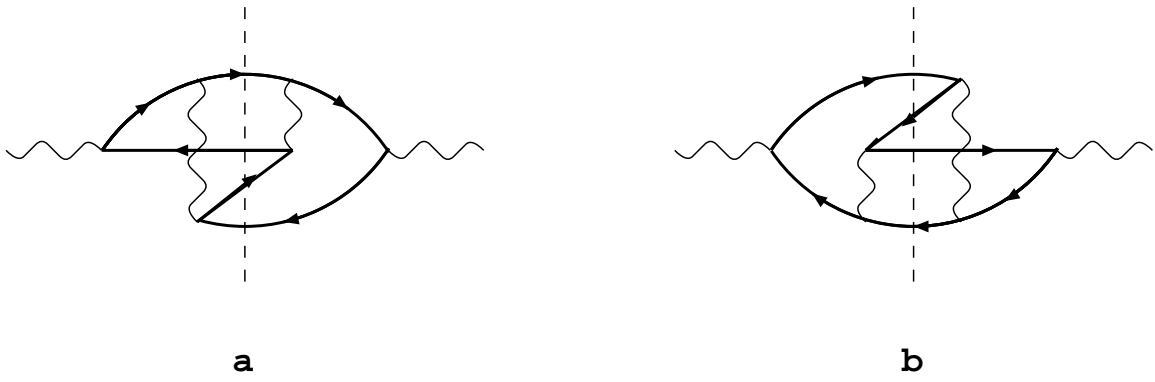
FIG. 6. The diagram Fig.5c represented in the Goldstone diagram technique.\\
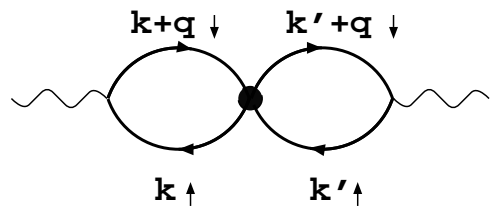
FIG. 7. Transform of the diagram Fig.6a into diagram with
effective ``dot'' interaction.\\
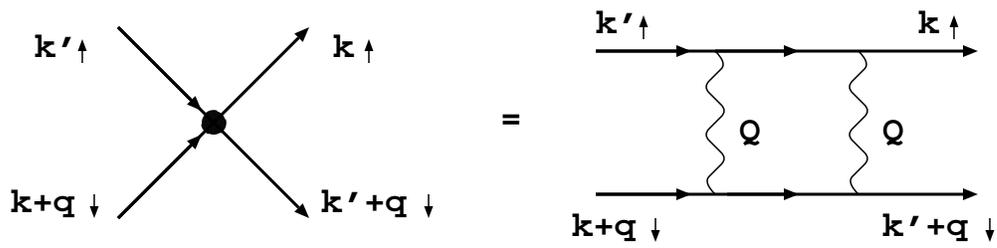
FIG. 8. Effective ``dot''.\\
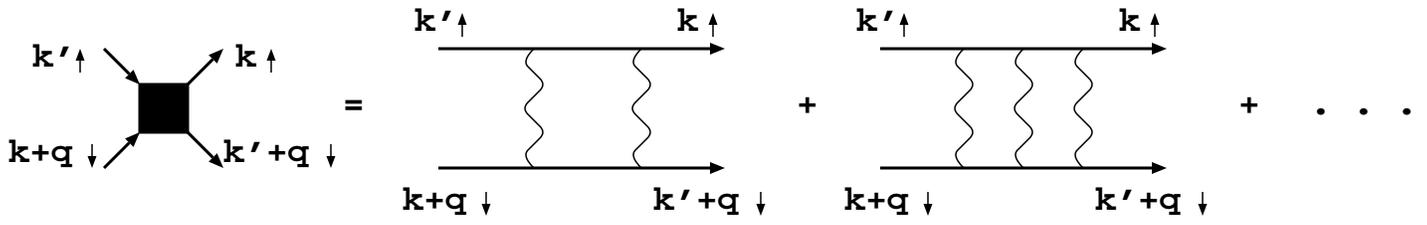
FIG. 9. The effective particle-hole interaction in the spin-flip channel.\\
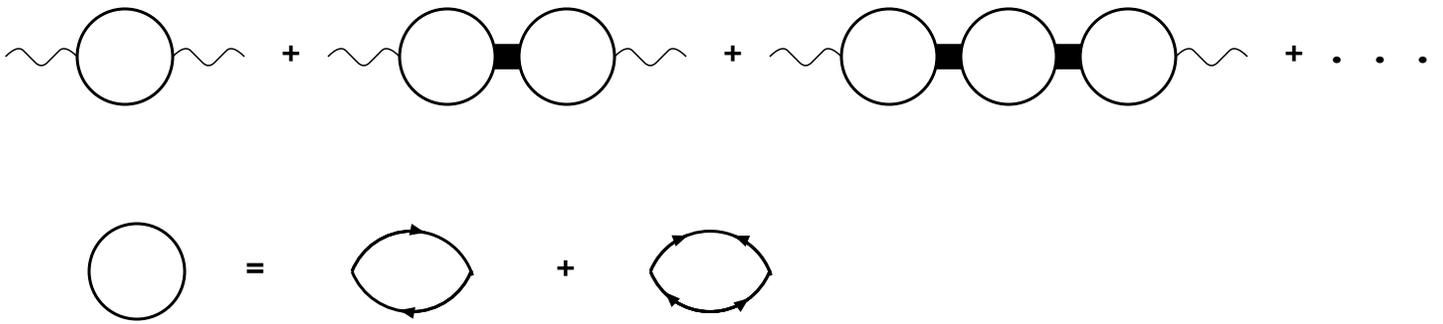
FIG. 10. Spin-wave polarization operator with chaining of
effective particle hole interaction taken into account.\\
FIG. 11. Effective spin-wave frequency $\Omega_{\nu{\bf q}}$ as a
function of the bare frequency $\omega_{\bf q}$.\\
Solid line: the direction ${\bf q}\propto (1,0)$.
Dashed line: the direction ${\bf q}\propto (1,1)$.
For comparison we present unrenormalized frequency
($\Omega_{\nu{\bf q}}$=$\omega_{\bf q}$): dashed-dotted line.\\
{\bf a)}: hole concentration $\delta=0.05$.\\
{\bf b)}: $\delta=0.1$. For this concentration we present also
the frequency of collective excitation $o_{\bf q}$ for the
direction ${\bf q}\propto (1,0)$.\\
{\bf c)}: $\delta=0.15$.\\
{\bf d)}: $\delta=0.2$.\\
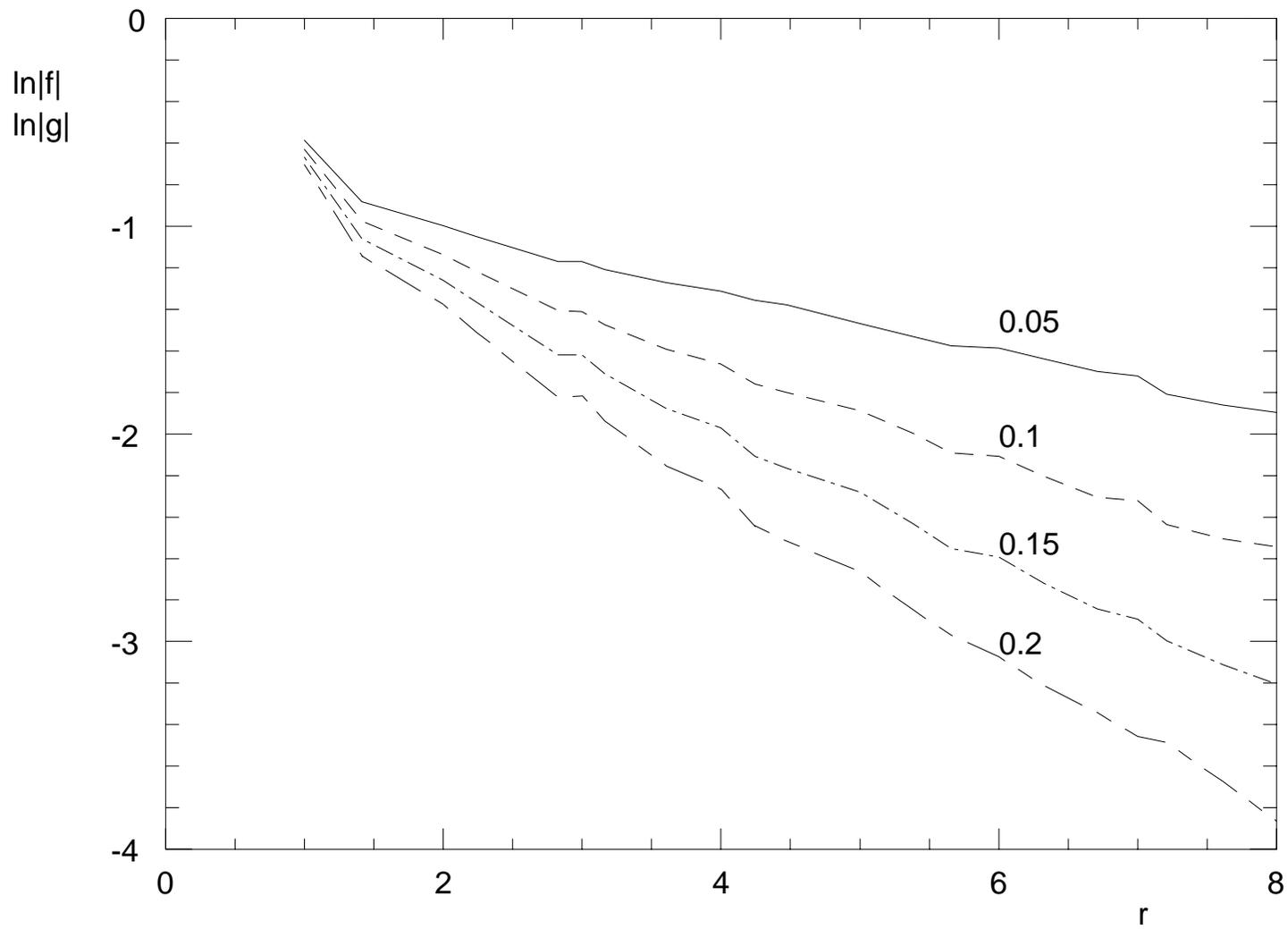
FIG. 12. $\ln |g|$ and $\ln |f|$ as a functions of the distance
between sites $r=\sqrt{m^2+n^2}$. If $m+n$ is odd, the $\ln{|g|}$
is plotted. And if $m+n$ is even, the $\ln{|f|}$ is plotted.
The curves correspond to hole concentrations $\delta=0.05,0.1,0.15,0.2$.\\
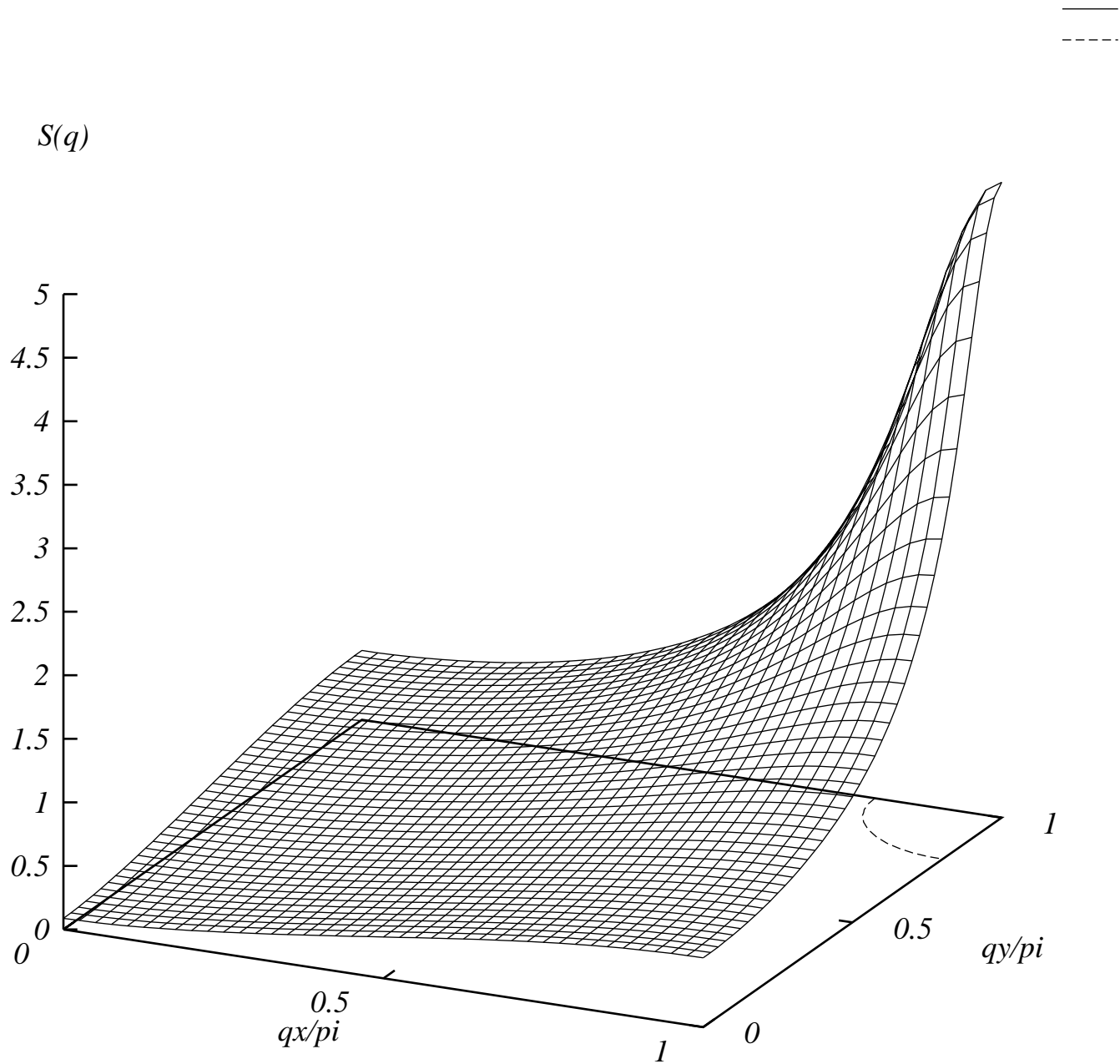
FIG. 13. Static magnetic formfactor $S_M({\bf q})$ at hole concentration
$\delta=0.15$. The arguments are $q_x/\pi$ and $q_y/\pi$.
The dashed contour corresponds to the half width.


\begin{references}
\bibitem{hole} Sometimes we use term ``hole'' in sense of $t-J$ model
and sometimes (usually in combination particle-hole) in sense of
Green's function technique. In each particular case the meaning of term is
evident from context and can not cause misunderstanding.
\end{references}
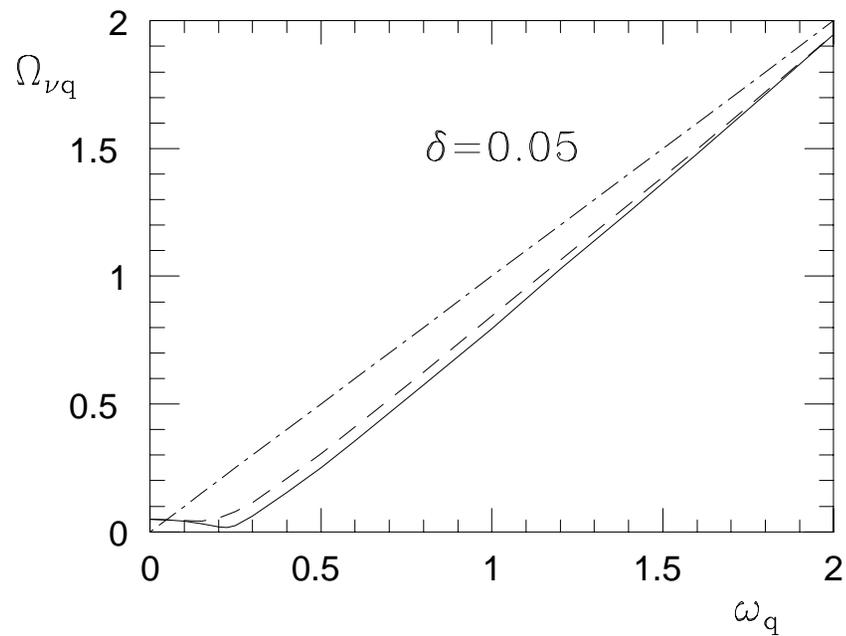
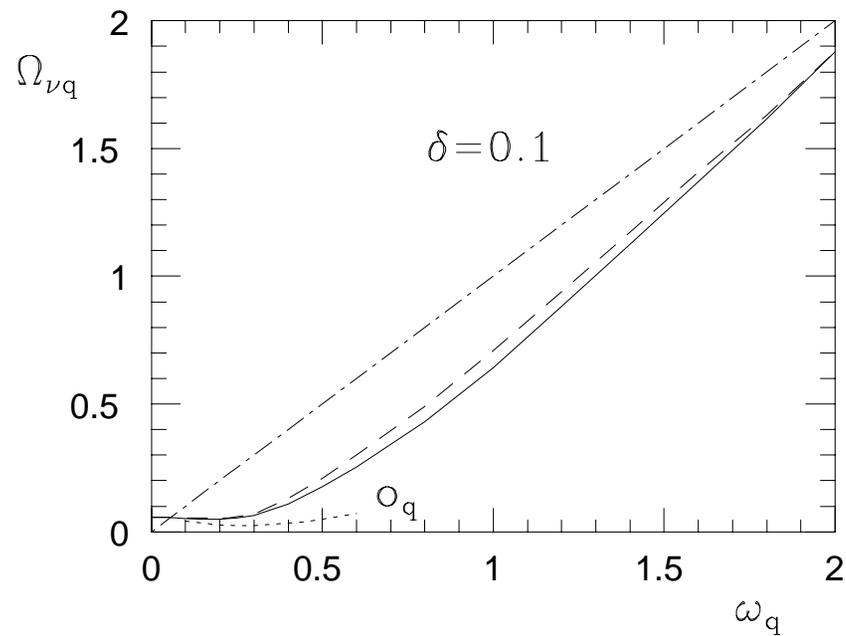
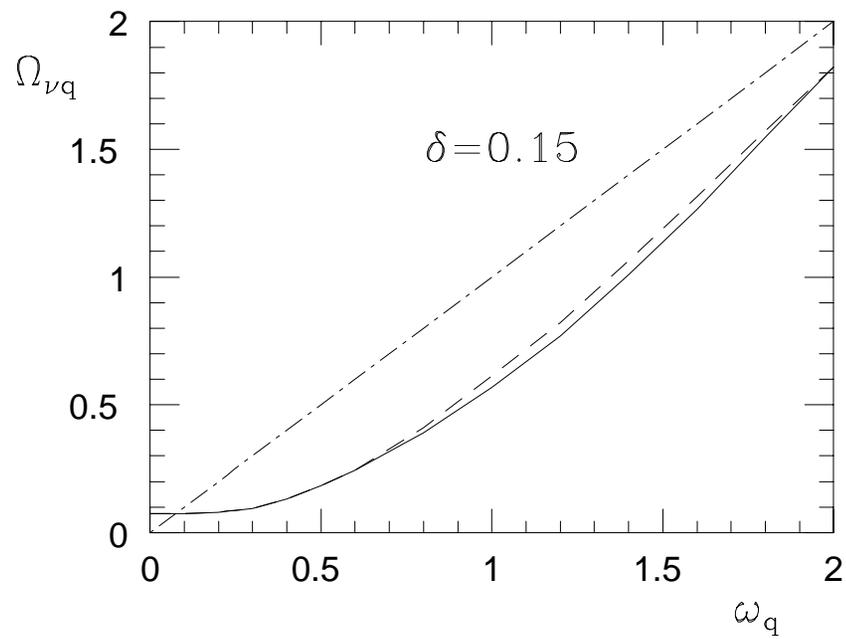
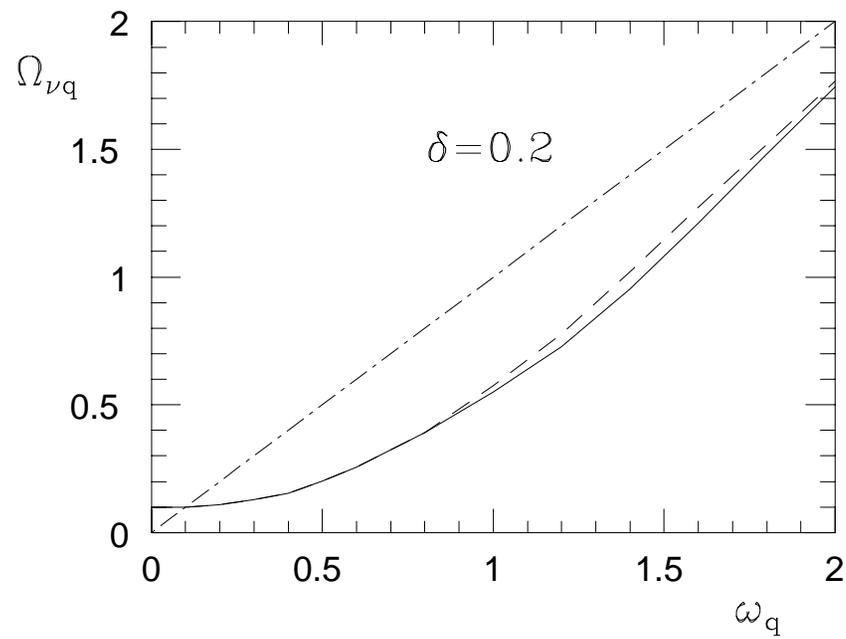
\end{document}